\documentclass[%
 aip,
 amsmath,amssymb,
 reprint,%
]{revtex4-1}

\usepackage{graphicx}
\usepackage{dcolumn}
\usepackage{bm}
\usepackage{xcolor}
\usepackage[utf8]{inputenc}
\usepackage[T1]{fontenc}
\usepackage{mathptmx}

\begin{document}

\preprint{AIP/123-QED}

\title{Griffiths-like phase close to the Mott transition}

\author{Isys F. Mello}
\affiliation{S\~ao Paulo State University (Unesp), IGCE - Physics Department, 13506-900, Rio Claro - SP, Brazil}
\author{Lucas Squillante}%
\affiliation{S\~ao Paulo State University (Unesp), IGCE - Physics Department, 13506-900, Rio Claro - SP, Brazil}%
\author{Gabriel O. Gomes}
\affiliation{University of S\~ao Paulo, Department of Astronomy, 05508-090, S\~ao Paulo, Brazil}%
\author{Antonio C. Seridonio}
\affiliation{S\~ao Paulo State University (Unesp), Department of Physics and Chemistry, 15385-000, Ilha Solteira - SP, Brazil}
\author{Mariano de Souza}
 \email{mariano.souza@unesp.br}
\affiliation{S\~ao Paulo State University (Unesp), IGCE - Physics Department, 13506-900, Rio Claro - SP, Brazil}

\date{\today}

\begin{abstract}
We explore the coexistence region in the vicinity of the Mott critical end point employing a compressible cell spin-$1/2$ Ising-like model. We analyze the case for the spin-liquid candidate $\kappa$-(BEDT-TTF)$_2$Cu$_2$(CN)$_3$, where close to the Mott critical end point metallic puddles coexist with an insulating ferroelectric phase. Our results are fourfold: \emph{i}) a universal divergent-like behavior of the Gr\"uneisen parameter
upon crossing the
first-order transition line; \emph{ii}) based on scaling arguments,
we show that within the coexistence region, for \emph{any} system close to the critical point, the relaxation time is entropy-dependent; \emph{iii}) we propose the electric Gr\"uneisen parameter $\Gamma_E$, which quantifies the electrocaloric effect;  \emph{iv}) we identify the metallic/insulating coexistence region as an electronic Griffiths-like phase. Our findings suggest that $\Gamma_E$ governs the dielectric response close to the critical point and that an electronic Griffiths-like phase emerges in the coexistence region.
\end{abstract}

\maketitle

%
\section{Introduction}
\begin{figure}
\centering
\includegraphics[clip,width=0.83\columnwidth]{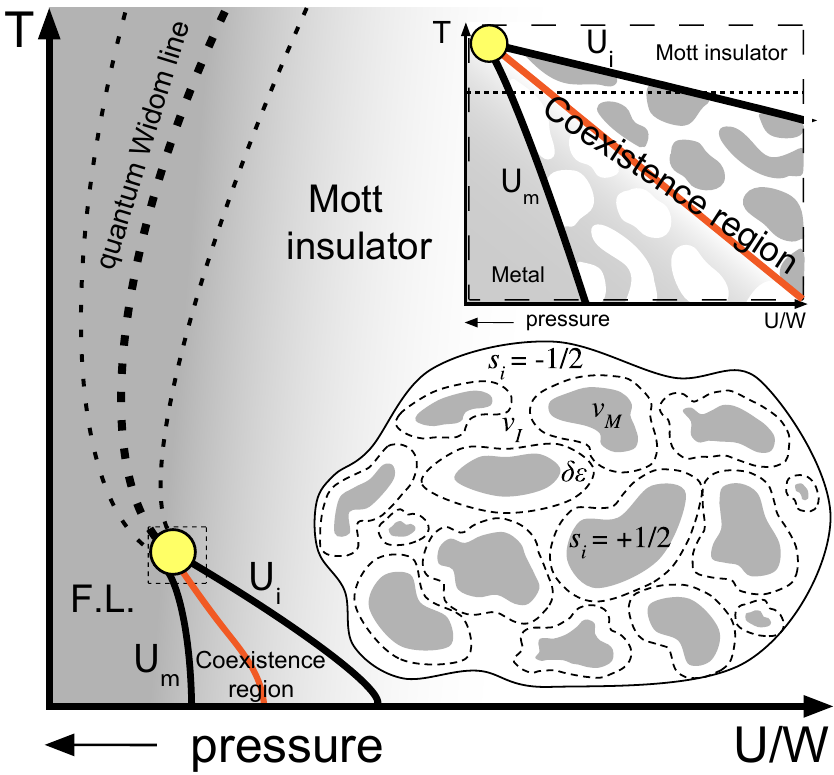}
\caption{Main panel: schematic temperature $T$ \emph{versus} $U/W \propto p^{-1}$  phase diagram of $\kappa$-(BEDT-TTF)$_2$Cu$_2$(CN)$_3$. The various phases, coexistence region, the quantum Widom line, as well as the critical point (yellow bullet) are indicated. F.L. refers to Fermi-liquid \cite{Dressel}. The metallic ($U_m$) and insulating ($U_i$) spinodal lines are depicted \cite{Jerome2009}, which  delimitates the phases coexistence region. The solid orange color line represents the first-order transition line. Upper inset: zoomed coexistence region with a cartoon of the metallic puddles embedded in the insulating phase. The horizontal dashed line represents the hypothetical $U/W$ sweep at constant temperature, namely $T/W$. Picture after Refs.\,\cite{vlad,vlad2}. Lower inset: cartoon of the here proposed electronic Griffiths-like phase, being metallic puddles ($s_i = +1/2$) with volume $v_M$ embedded in an insulating medium ($s_i = -1/2$) with volume $v_I$. The dashed lines represent the interaction $\delta\varepsilon$ between neighbouring metallic puddles.}
\label{Fig-1}
\end{figure}

Over the last few decades the Mott metal-to-insulator (MI) transition has been investigated intensively, see, e.g., Refs.\,\cite{Imada,Barto,Souza2015}. Indeed, the Mott MI transition constitutes the paradigm in the field of strongly correlated electronic phenomena.
The ratio of the on-site Coulomb repulsion $U$ to the bandwidth $W$, i.e.\,, $U/W$, is the so-called Hubbard parameter \cite{Imada}. It turns out that $U/W \propto p^{-1}$, where $p$ refers to the applied pressure, constitutes the tuning parameter of the Mott MI transition.
Much efforts have been made to understand the emergence of a superconducting phase from the Mott insulating dome upon pressurising the system, as it occurs, for instance, in molecular conductors and cuprates \cite{Dagotto2005}.
As can be seen in the schematic phase diagram depicted in the main panel of Fig.\,\ref{Fig-1}, upon applying pressure $U/W$ is reduced and  a Mott transition takes place. Particular attention has been paid to unveil the Physics on the verge of the Mott second-order critical end point, which is marked by maxima in response functions \cite{Gene,PRL2007,Fournier}. Recently, the phases coexistence close to the Mott critical point of $\kappa$-(BEDT-TTF)$_2$Cu$_2$(CN)$_3$  was investigated using dielectric spectroscopy \cite{Dressel}. Upon approaching the critical parameters, the quasi-static dielectric constant $\varepsilon_1$ is dramatically enhanced, achieving  $\varepsilon_1 \approx$ 10$^5$. Considering pressure-induced critical points, the so-called Gr\"uneisen parameter has been broadly employed \cite{EPJ,Zhu,Barto,Souza2015}.
Experimentally, the insulating and metallic puddles can, respectively, be probed by standard dielectric \cite{Dressel} and resistance measurements \cite{kanoda2}. Here, we employ a Thermodynamics-based $S = 1/2$ Ising-like model \cite{Cerdeirina2019,SR} to investigate the coexistence region close to the Mott second-order critical end point. The here proposed dielectric Gr\"uneisen parameter $\Gamma_{\varepsilon}$ enables us to describe in a 
way the dielectric response in the coexistence region in terms of the Hubbard parameter. 
\section{The Model}
The well-known magnetic Ising model considers the possible spin configurations and the magnetic coupling between neighboring spins \cite{Ising1925}. Analogously, the Ising model can be employed in the investigation of several other scenarios, such as in econophysics \cite{Schinckus}, democratic elections \cite{Siegenfeld}, and even in the spread of diseases \cite{Crisostomo2012, epidemics}. In this context, another example is the application of a spin $S$ = 1 Ising model for the $\lambda$-transition of the mixture composed by He$^3$ and He$^4$\,\cite{Griffiths1971}. In the present work, we have employed the so-called compressible Ising model \cite{Cerdeirina2019}, which essentially takes into account the coexistence of two accessible volumes in the system. The main idea behind the model is to associate the up and down spin configurations in the magnetic Ising model with two coexisting accessible volumes into the system of interest. Recently, the compressible Ising model was employed for the supercooled phase of water \cite{SR}, where the two coexisting accessible volumes considered were the high- and low-density phases. In order to describe the coexistence region close to the Mott critical end point, we employ a compressible cell Ising-like model \cite{Cerdeirina2019, SR}. An usual approach to incorporate disorder effects is based on the Random Field Ising Model (RFIM) \cite{Carlson2006, Doria2015}, where the local field associated with the different regions in the system is taken into account. However, since our focus lies on the understanding of the coexistence region employing the metallic volume fraction obtained via the Bruggeman equation in Ref.\,\cite{Dressel}, we present an alternative approach to treat spatial random disorder effects. We make use of the metallic volume fraction $x = \frac{1}{2}\tanh{\left[c\frac{(U/W)_{crit}- (U/W)}{(T/W)_{crit} - (T/W)}\right]} + \frac{1}{2}$, reported in Ref.\,\cite{Dressel},  where $(U/W)_{crit} = (0.20 - T/W)/0.14$ and $(T/W)_{crit}$ represent, respectively, the values of $U/W$ and $T/W$ at the critical point, and $c$ is a non-universal constant. It is clear that we have to analyze the various quantities as a function of $x$, since $x$ can be seen as \emph{pseudo} tuning parameter. Experimentally, application of pressure affects the $U/W$ ratio, which in turn reflects in $x$. In our approach, we consider an insulating matrix with volume $v_I$, within which there is place to the emergence of metallic bubbles upon pressurization of the system, so that $x = v_M/v_T$, where $v_M$ and $v_T$ are the metallic and total volume, respectively. Note that for $x \rightarrow 1$, $v_M \rightarrow v_T$, whereas for $x \rightarrow 0$, $v_M \rightarrow 0$. Thus, $v_T = N^Mv_M + N^Iv_I$, where $N^M$ and $N^I$ refers to the number of metallic and insulating puddles, respectively. We consider a $S = 1/2$ Ising-like model,  so that we associate metallic (insulating) puddles with $S = + 1/2$ ($S = - 1/2$), see lower inset of Fig.\,\ref{Fig-1}. Yet, to account  $N^M$ and $N^I$ properly, we define  $N^{M} = \sum_{i=1}^{N}2(s_i+1/2)s_i$ and $N^{I} = \sum_{i=1}^{N}2(s_i-1/2)s_i$, where $s_i$ is the  \emph{label} of a single puddle and $N$ the total number of puddles, i.e., $N  = N^M + N^I$.
The total energy $E^T(s_i)$ can be written taking into account the term related to the energy $\delta\varepsilon '$ associated with each metallic puddle, being the energy $\varepsilon_0$ associated with the insulating puddle:
\begin{equation}
E^T(s_i) = N\varepsilon_0 - \delta\varepsilon'\sum_{<i \neq j>}^{N}\left[2\left(s_i+\frac{1}{2}\right)s_i\right]\times\left[2\left(s_j+\frac{1}{2}\right)s_j\right],
\end{equation}
where $i$, $j$ label neighbouring puddles and $N\delta\varepsilon' = \delta\varepsilon$.
Considering that we are dealing with a mixture composed by metallic and insulating phases, we employ the Gibbs free energy of mixing $\Delta g$ associated with the mixture of phases and the configuration of the metallic puddles, which reads \cite{Atkins, Mitropoulos}:
\begin{equation}
\Delta g = [x(1-x)\delta\varepsilon] + k_BT[x\ln{(x)} + (1-x)\ln{(1-x)}],
\label{Gibbs}
\end{equation}
where $k_B$ is Boltzmann constant and $\delta\varepsilon$ emulates the interaction between neighbouring metallic puddles.
In analogy with the Hubbard model,  $\delta \varepsilon$  could be associated with the hopping term $t$, so that we have for the insulating phase $\delta \varepsilon = 0$, whereas for the metallic phase $\delta \varepsilon \neq$ 0.  As $U/W$ is reduced, $W \propto t  \propto \delta \varepsilon$  is increased and thus metallic puddles emerge. The first term of Eq.\,\ref{Gibbs} is associated with the metallization enthalpy, while the second one is related to the configurational entropy.
The total entropy $s$ of the system is the sum of the configurational entropy $s_{conf}$ and the intrinsic entropy  $s_{part}$ within the puddles  \cite{Cerdeirina2019}. Hence,
\begin{eqnarray}
\nonumber s(x, T) &=& k_B\left[-x \ln (x)-(1-x) \ln(1-x)\right]+\frac{1}{2}k_B \times \\
&& \left\{ \ln \left[ \frac{v_M v_I h^{-6}}{(2\pi m k_B T)^{-3}} \right] + (2x-1)\ln \left( \frac{v_M}{v_I} \right)\right\},
\label{entropy}
\end{eqnarray}
where $m$ is the mass of the set of particles confined in the puddle and $h$ is Planck's constant.
From Eq.\,\ref{entropy}, it is straightforward to derive the observables of interest. Using Maxwell-relations \cite{Gene}, $\frac{p}{T}=\left(\frac{\partial s}{\partial v_M}\right)_{N,E}$, we obtain 
 the equation of state for the coexistence region:
\begin{eqnarray}
p(T, x)&=& k_B T \{v_T^{-1}\ln(1-x) - v_T^{-1}\ln(x) + \frac{1}{2} \times \nonumber \\
&& \times [ -(v_Tx)^{-1}(2x-1)(v_T-v_Tx)A +\nonumber \\
&& + (v_T-2v_Tx)
(v_Tx(v_T-v_Tx))^{-1} - \nonumber \\ && - 2 (v_T)^{-1} \ln[(1-x)^{-1}] ]     \}
\end{eqnarray}
where $A=[v_Tx/(v_T-v_Tx)^2+ 1/(v_T-v_Tx)]$.
The Gr\"uneisen parameter can be obtained using the relation $\Gamma = \frac{1}{V_m T}\left( \frac{\partial T}{\partial p} \right)_s$ \cite{EPJ}, where $V_m$ is the molar volume,  which in turn enables us to determine $\Gamma_{\varepsilon}$.
\begin{figure}[h!]
\centering
\includegraphics[clip,width=0.93\columnwidth]{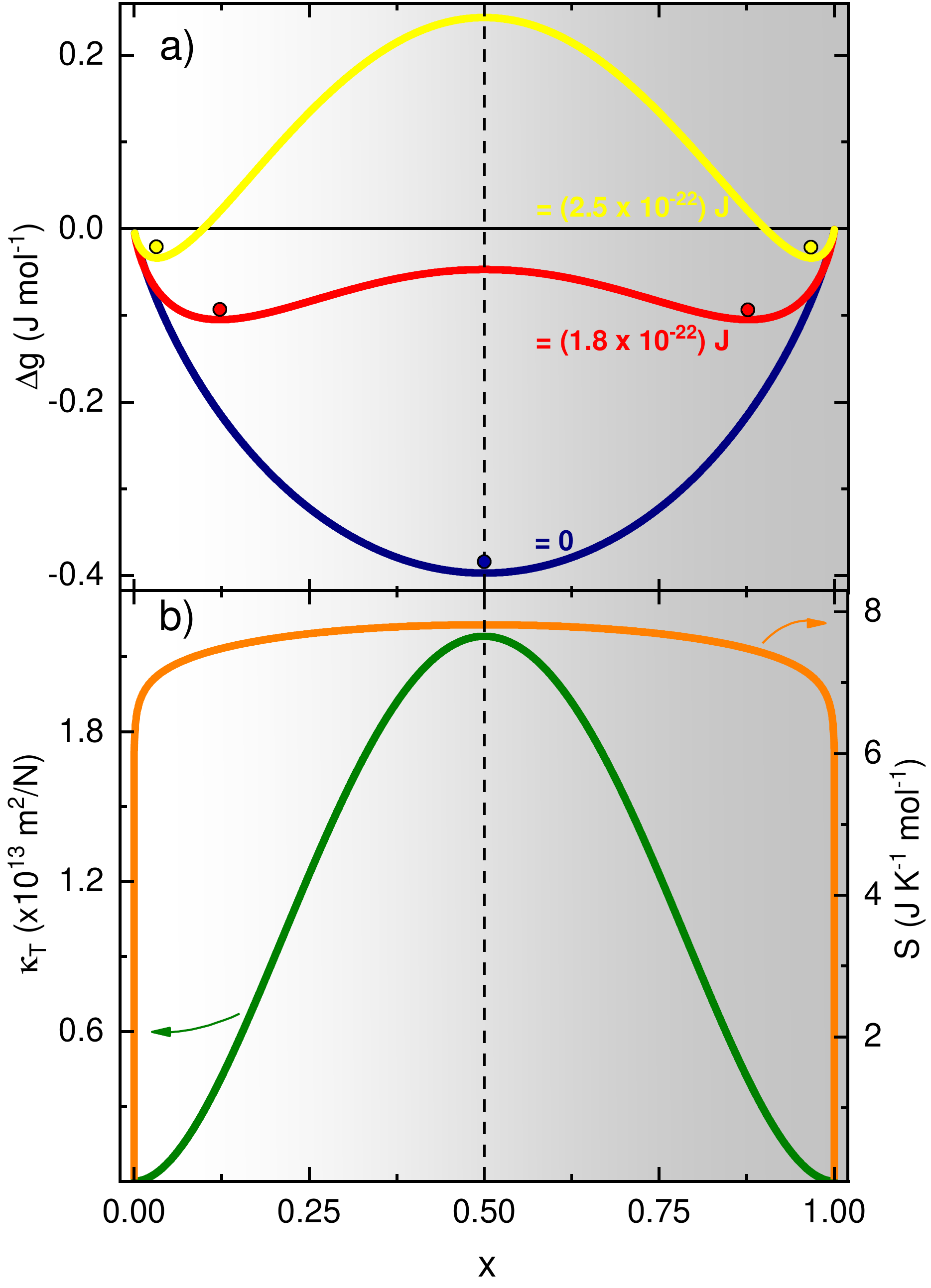}
\caption{ a) The Gibbs free energy $\Delta g$ \emph{versus} $x$ for different interaction energy $\delta \varepsilon$ values. The bullets indicate the various minima \cite{Mitropoulos}.   b) Isothermal compressibility $\kappa_T$ (green line) and entropy $S$ (orange color line) \emph{versus} $x$. The vertical dashed line is centered at $x =$ 0.5.}
\label{Fig-2}
\end{figure}
\begin{figure}[h!]
\centering
\includegraphics[clip,width=0.95\columnwidth]{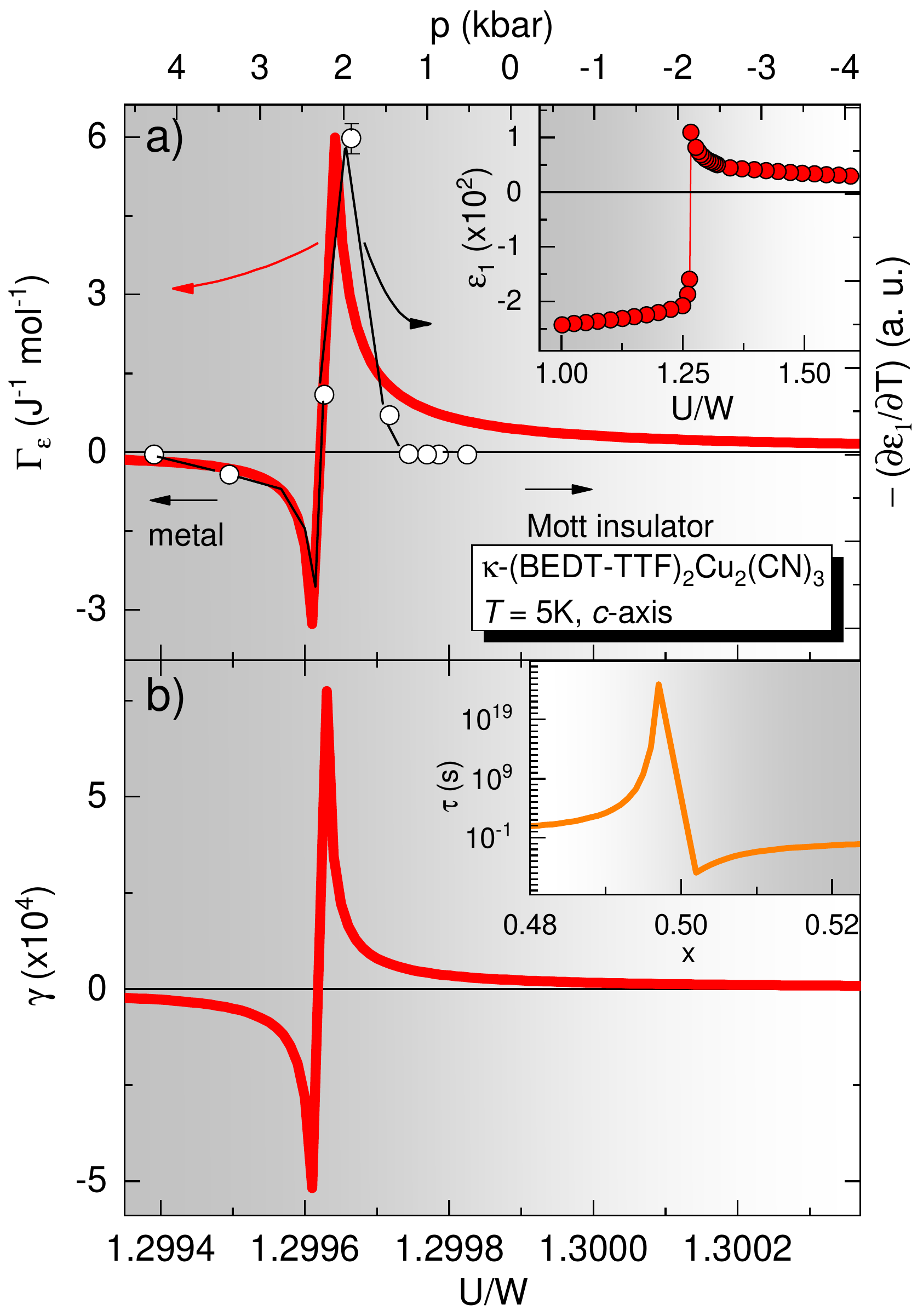}
\caption{ a) Main panel: the red line (left and lower axis) is the dielectric Gr\"uneisen parameter $\Gamma_{\varepsilon}$ as a function of $U/W$ for $\kappa$-(BEDT-TTF)$_2$Cu$_2$(CN)$_3$  at $T = 5\,$K along the crystallographic $c$-axis, cf.\,label. We have used $W = (3.8221 \times 10^{-21})$\,J \cite{Pustogow}. White circles (upper and right axis) refers to experimental data, namely the temperature derivative of the dielectric constant $\varepsilon_1$ \emph{versus} pressure $p$ at $T =$ 20\,K, data set taken from Ref.\,\cite{pustgow2}. The solid line is guide for the eyes. Inset: dielectric constant $\varepsilon_1$ \emph{versus} $U/W$, obtained from dynamical mean-field theory, data taken from Ref.\,\cite{Pustogow}.  b) Main panel: Gr\"uneisen parameter $\gamma$ as a function of $U/W$. $\gamma$ \emph{versus} $U/W$ was obtained using $v_0 = (2.4164317\times10^{-8})\,\textmd{m}^3$, $q = 9.148786\times10^{-9}$, $U/{W_0} = 0.89$, $C = 77.37$, $\tau = (0.2\times10^{-5})\,\textmd{s}$, and $\tau_0 = (2.5\times10^{-8})\,\textmd{s}$. Inset: the relaxation time $\tau$ \emph{versus} metallic volume fraction $x$.}
\label{Fig-3}
\end{figure}
\section{Results and Discussion}
Figure \ref{Fig-2} a) depicts the free energy $\Delta g$ $versus$ the metallic volume fraction $x$. Note that in the absence of interaction ($\delta \varepsilon$ = 0) between metallic puddles a single minimum centered in $x$ = 0.5 is observed. Upon increasing $\delta \varepsilon$, two symmetric  minima shows up close to the single phase boundaries. As can be seen in Fig.\ref{Fig-2} b), for $N^M = N^I$, i.e., $x = 0.5$, the entropy is maximized and the system becomes more compressible. Note that $\kappa_T$ is non-singular, since the system is close to the critical point.  Making use of the Hubbard parameter obtained via optical measurements, reported in Ref.\,\cite{Pustogow}, we compute $\Gamma_\varepsilon$ for $\kappa$-(BEDT-TTF)$_2$Cu$_2$(CN)$_3$, as depicted in Fig.\,\ref{Fig-3} a). Following discussions made in Ref.\,\cite{Dressel}, in the coexistence region, upon application of pressure, $\varepsilon_1$ increases  until the metallic fraction becomes higher than the insulating one. Upon further increasing pressure,  a sign change is observed and $\varepsilon_1$ assumes negative absolute values. Such a peculiar behavior is also observed in $\Gamma_\varepsilon$. Furthermore, $\Gamma_{\varepsilon}$ follows the same behaviour as $\varepsilon_{1}$ as a function of $U/W$ obtained via DMFT simulations in Ref.\,\cite{Dressel}, cf.\,inset of Fig.\,\ref{Fig-3}a). As can be seen in Fig.\,\ref{Fig-3} a), in the limit $U/W \rightarrow (U/W)_{crit}$, i.e., $x =$ 0.5, $\Gamma_{\varepsilon} \rightarrow$ 0. In other words, $\Gamma_{\varepsilon}$ is dramatically enhanced near $U/W$ values corresponding to the coexistence region between insulating and metallic phases.
Note that for \emph{any} system where $x$ and the microscopic parameters are known, the behavior of $\varepsilon_1$ can be predicted via $\Gamma_{\varepsilon}$. This is one of the key results of this work.
Having in mind that the response functions are maximized around $x$ = 0.5, the coexistence region of the Mott transition can be interpreted in the frame of a spinodal decomposition \cite{Mitropoulos,Debenedetti}, i.e., the emergence of the metallic puddles upon applying pressure can be understood similarly to a nucleation process. Recent $^{13}$C nuclear magnetic resonance measurements on x-ray irradiated charge-transfer salts detected the presence of extremely slow electronic fluctuations in the phase boundary of the Mott transition, which were recognized by the authors as an electronic Griffiths-like phase \cite{Yamamoto}. The slow dynamics of electronic excitations inherent to the coexistence region is also captured in our approach and will be discussed into more details below. Before doing so, we recall that the magneto-caloric effect is quantified by the magnetic Gr\"uneisen parameter \cite{prbmce}. Analogously, we propose here the electric Gr\"uneisen parameter $\Gamma_E$, which in turn quantifies the electrocaloric effect \cite{ECE}:
\begin{equation}
\Gamma_E =\frac{\left(-\frac{\partial P}{\partial T} \right)_E }{c_E},
\label{ECE}
\end{equation}
 where $c_E$ and $P$ refer, respectively, to the specific heat at constant electric field $E$ and the electric polarization. The electric polarization is given by $P = \varepsilon_0\chi E$, where $\varepsilon_0$ is the vacuum permittivity, $\chi$ is the electric susceptibility \cite{aschcroft}. Using $\chi = [(\varepsilon/\varepsilon_0) - 1]$ we have $P = \left(\varepsilon - \varepsilon_0\right)E$, so that we can write
$\left(\frac{\partial P}{\partial T}\right)_E = E\left(\frac{\partial \varepsilon}{\partial T}\right)_E$ and thus $\Gamma_{E} = \frac{E}{c_E}\left(-\frac{\partial \varepsilon}{\partial T}\right)_E$.
Hence, it becomes evident that an enhancement of $\left(\frac{\partial \varepsilon}{\partial T}\right)_E$ implies that $\Gamma_E$ is also increased upon approaching the
first-order transition line. Note that $\Gamma_E$ mirrors the behavior of $\varepsilon$ when transiting in the coexistence region \cite{pustgow2}.
For relaxor-type ferroelectrics, application of external pressure can dramatically alter the dielectric properties of the system. More specifically, the system becomes less polarizable upon pressurising it due to the increasing of the Coulomb repulsion, i.e., the entropy is enhanced and so does the strength of the applied electric field required to polarize the system. Following such arguments, we infer that, in this physical scenario, the dielectric relaxation time $\tau$ also increases upon pressurising the system. This means that in the coexistence region, close to the critical point, $\tau$ depends on the entropy.
Now, we focus our analysis on systems where the presence of random spatial disorder leads to a frequency-dependent activation energy. In the frame of the model proposed by Avramov \cite{Casalini2, Paluch2003, Avramov1988}, the so-called jump frequency $f_i$, i.e., the frequency in which a particle in a given region labeled by $i$ can overcome an energy barrier $E^{'}$ in systems presenting random spatial disorder is given by:
\begin{equation}
f_i (E^{'}_i)= f_{0} \exp\left(-\frac{E^{'}_i}{k_BT}\right),
\label{fi}
\end{equation}
where $f_0$ is a constant, and $E^{'}_i$ is the activation energy. In order to define the mean value $\langle f_i \rangle$ of the jump frequency one must consider the probability $\varphi(E^{'}, \sigma)$ of overcoming $E^{'}$, so that \cite{Casalini2}:
\begin{equation}
\langle f_i \rangle=\int_0^{E^{'}_{max}}f(E)\varphi(E^{'},\sigma)dE^{'},
\label{averagenu}
\end{equation}
where $E^{'}_{max}$ refers to the maximum value of $E^{'}$ and $\sigma$ is the dispersion of the energies, which is related to the degree of disorder inherent to the investigated system. The parameter $\sigma$ is defined in terms of $S$, $k_B$, the coordination number $Z$, and the reference parameters, such as the entropy $S_r$ and the dispersion $\sigma_r$. The expression for $\sigma$ reads \cite{Casalini2}:
\begin{equation*}
\sigma = \sigma_r\exp{\left[\frac{2(S-S_r)}{Zk_B}\right]}.
\end{equation*}
Following discussions from Ref.\,\cite{Avramov1988}, we choose $\varphi$ as the Poisson distribution \cite{Casalini2}:
\begin{equation}
\varphi(E^{'},\sigma)=\frac{\exp[\frac{(E^{'}-E^{'}_{max})}{\sigma}]}{\sigma[1-\exp(-\frac{E^{'}_{max}}{\sigma})]},
\label{Poisson}
\end{equation}
Replacing Eqs.\,\ref{fi} and \ref{Poisson} into Eq.\,\ref{averagenu}, we obtain:
\begin{equation*}
\langle f_i\rangle = \int_0^{E^{'}_{max}}f_{0} \exp\left(-\frac{E^{'}_i}{k_BT}\right)\frac{\exp[\frac{(E^{'}-E^{'}_{max})}{\sigma}]}{\sigma[1-\exp(-\frac{E^{'}_{max}}{\sigma})]}dE^{'}
\end{equation*}
\begin{equation*}
\langle f_i\rangle = f_0 \frac{k_BT}{(k_BT-\sigma)}\frac{\exp\left[ \frac{(k_BTE^{'}_{max}-\sigma E^{'}_{max})}{\sigma k_BT} \right]-1}{\exp\left[ \frac{E^{'}_{max}}{\sigma}  \right]-1}
\end{equation*}
\begin{equation}
\langle f\rangle \simeq f_0 \exp\left(\frac{-E^{'}_{max}}{\sigma}\right),
\label{averagefrequency}
\end{equation}
and thus, replacing $\sigma$ in Eq.\,\ref{averagefrequency}, $f$ reads:
\begin{equation}
f = f_0 \exp\left\{-\frac{E^{'}_{max}}{\sigma_r}\exp\left[-\frac{2(S-S_r)}{Zk_B} \right]\right\}.
\label{frequency}
\end{equation}
Knowing that the frequency is inversely proportional to the relaxation time $\tau$ by the elementary relation $f = 1/\tau$, Eq.\,\ref{frequency} can be rewritten in terms of $\tau$  \cite{Casalini2}:
\begin{equation}
\tau=\tau_0\exp\left\{\frac{E^{'}_{max}}{\sigma_r}\exp\left[-\frac{2(S-S_r)}{Zk_B} \right]\right\},
\label{relaxation}
\end{equation}
where $\tau_0$ is a constant. Also, since the entropy is a function of $T$ and $v$, we have \cite{Slater_book}:
\begin{equation}
dS=\left(\frac{\partial S}{\partial T} \right)_v dT + \left(\frac{\partial S}{\partial v} \right)_T dv,
\end{equation}
which can be rewritten as \cite{Casalini2}:
\begin{equation}
dS=\frac{c_v}{T}dT + \frac{(c_p-c_v)}{v\alpha T}dv.
\label{equationfords}
\end{equation}
Integrating both sides of Eq.\,\ref{equationfords} and taking into account the reference values, we have \cite{Casalini2}:
\begin{equation}
(S-S_r)= c_v \left[\ln\left(\frac{T}{T_r}\right) + \frac{(c_p-c_v)/c_v}{\alpha T}\ln\left(\frac{v}{v_r}\right)\right].
\label{entropy_dif}
\end{equation}
Replacing $\gamma=\frac{(c_p-c_v)/c_v}{\alpha T}$ in Eq.\,\ref{entropy_dif}:
\begin{equation}
(S-S_r)= c_v \left[\ln\left(\frac{T}{T_r}\right) + \gamma\ln\left(\frac{v}{v_r}\right)\right].
\label{s-sr}
\end{equation}
Finally, replacing Eq.\,\ref{s-sr} in \ref{relaxation} we achieve the key mathematical expression used in our analysis \cite{Casalini2}:
\begin{equation}
\tau=\tau_0\exp\left[\frac{E^{'}_{max}}{\sigma_r}\left( \frac{T_r v_r^{\gamma}}{Tv^{\gamma}} \right)^{\frac{2c_v}{Zk_B}}\right] .
\label{relaxation_gamma}
\end{equation}
Equation \ref{relaxation_gamma} establishes clearly the connection between $\tau$ and $\gamma$ and enables us to explore the relaxation time for systems with random spatial disorder when the thermodynamic quantities are known.
Thus, for any system with entropy-dependent $\tau$, the volume $v$ scaling exponent is the Gr\"uneisen parameter $\gamma$ itself (we make use of $\gamma$ to refer to the Gr\"uneisen parameter obtained from scaling arguments) \cite{Casalini2}, namely $\tau(T,v,\gamma) = \tau_0\textmd{exp}\left(\frac{C}{Tv^{\gamma}}\right)$, where $\tau_0$ is a typical time scale of an a.c.\,response at high-temperatures and $C$ is a non-universal constant.
It is worth mentioning that the relaxation time of the phases coexistence region of the Mott transition is associated with the intrinsic random spatial disorder, which in turn is related to the presence of metallic puddles embedded in the insulating matrix. Note that in the presence of random spatial disorder the time-scale needed to reverse the electric polarization of the system is enhanced. Since the presence of spatial disorder leads to an entropy increase, it becomes evident that $\tau$ is entropy dependent. Usually, the well-known Vogel-Fulcher law is employed to describe the dependence of the ferroelectric transition temperature in terms of the frequency \cite{abdel}, considering a single activation energy, i.e., a single frequency associated with the dielectric response of the system. However, such a model does not incorporate the entropy variation and its influence on $\tau$ \cite{Casalini2}. As a matter of fact, a proper description of the entropy dependence of the relaxation time can be performed through the Avramov model here employed \cite{Casalini2}. This is corroborated by the fact that such a model assumes a distribution of distinct activation energies in the system and, as a consequence, distinct relaxation times. In our approach, using the Avramov’s model and the connection between the Gr\"uneisen parameter and the relaxation time for systems with entropy-dependent $\tau$, we employed an expression for $\tau$ which incorporates random intrinsic disorder \cite{Casalini2}. Also, it is worth mentioning that the spin-liquid candidate system here discussed presents an intrinsically glass-like response in the dielectric constant \cite{abdel, Sakai2014}. This behavior can be understood in terms of the presence of a magnetic frustration inherent to this system, since there is a strong coupling between the electric polarization and the spin degrees of freedom \cite{Sakai2014}. Thus, the intrinsic disorder associated with the proposed glass-type behavior impacts the relaxation-time of the system, making the dielectric constant to be frequency-dependent even under applied pressure \cite{abdel}. The application of pressure, in turn, makes metallic puddles (non-polarized regions) to appear within the insulating matrix, which significantly increases the entropy leading thus to a slowing down of the system upon crossing the first-order transition line \cite{Casalini2}.
It becomes then clear that the volume dictates the $T$ and $p$ dependence of $\tau$. It is thus straightforward to write $\gamma(T,v,\tau) = \frac{\log{\left[\frac{C}{T\log{(\tau/\tau_0)}}\right]}}{\log{(v)}}$.
In the tight-binding approximation $W(v) = W_0\textmd{exp}\left[-q\left(\frac{v - v_0}{v_0}\right)\right]$, where $v_0$ and $W_0$ are, respectively, the volume and bandwidth at ambient pressure and $q$ is the volume dependence of the overlap integral \cite{Cyrot}. Thus, we can write $\gamma$, as follows:
\begin{equation}
\gamma[\tau,(U/W),T] = \frac{\log{\left[\frac{C}{T\log{(\tau/\tau_0)}}\right]}}{\log{\left[v_0+\frac{q(\log{(U/W_0)+\log{(U/W)}})}{v_0}\right]}}.
\end{equation}
Hence, following scaling arguments \cite{Casalini2}, our analysis suggests that in the proximity of any finite-$T$ critical end point $\tau$ is entropy dependent.
Thus, $\tau$ can be linked with the Gr\"uneisen parameter using the scaling-relation $\tau (T,V,\gamma)\propto (C/TV^{\gamma})$.
Just to mention, for spin-glass systems the so-called time-field scaling is well-established, which is associated with the slow-dynamics regime of the system and it is given by $G_z(t, H) = G_z(t, H^{\gamma'})$ \cite{Keren1996, Tripathi2019}, where $t$ is the time, $\gamma'$ is the scaling exponent, and $H$ is the external magnetic field. The similarity between Eq.\,\ref{relaxation_gamma} derived in the frame of the Avramov model and $G_z(t, H)$ is remarkable. Such a similarity shows up because in both cases the glass-type behavior is contemplated \cite{Casalini2}. In the present case, a direct analogy with spin-glass systems can be performed. This is because the polarized regions are independent of each other, each one of them has a distinct electric polarization orientation.
Note that for $x = 0.5$, both $\gamma$ and $\Gamma_{\varepsilon}$ are zero [Fig.\,\ref{Fig-3}a)] making thus $\tau$ to be maximized, cf. Fig.\,\ref{Fig-3}b). Indeed, it turns out that upon approaching the first-order transition line, namely close to $x$ = 0.5, $\tau$ is significantly enhanced, slowing down the dynamics of the electronic fluctuations which are detected experimentally via low-frequency dielectric response measurements \cite{abdel}. This is in perfect agreement with the quasi-static dielectric constant experiments under pressure for various frequencies reported in Ref.\,\cite{Dressel}, where the maximum dielectric response for $\kappa$-(BEDT-TTF)$_2$Cu$_2$(CN)$_3$ is observed at low-frequencies for pressure values close to the critical one.
Making use of literature values for the various parameters \cite{abdel,Dressel,Cyrot}, the behavior of $\gamma$ as a function of $x$ can be analyzed, cf.\,shown in the main panel of Fig.\,\ref{Fig-3} b). It turns out that $\gamma$ also presents a divergent-like behavior near the critical values of $U/W$ in a similar way of $\Gamma_{\varepsilon}$, in agreement with the scaling proposal \cite{Casalini2}. Such findings support the existence of entropy-dependent relaxation process. Also, $\tau \rightarrow \infty$ for $x = 0.5$, indicating thus that the system lies in a slow dynamics regime, corroborating the spinodal decomposition analysis \cite{Mitropoulos}. Note that $\Gamma_{\varepsilon}$ is only valid within the coexistence region, while the relation between $\gamma$ and $\tau$ is universal \cite{Casalini2}. By employing  $\Gamma_{\varepsilon}$, $\Gamma_E$ and $\gamma$, the dielectric properties of relaxor-type ferroelectrics under pressure/doping can not only be described, but also predicted close to critical points. Furthermore, considering that $\Gamma = -\frac{1}{TV_m}\frac{\left(\frac{\partial S}{\partial p}\right)_T}{\left(\frac{\partial S}{\partial T}\right)_p}$, our findings suggest that in the vicinity of \emph{any} critical point, $\tau$ is entropy-dependent. Indeed, the singularities in both $\Gamma_{\varepsilon}$ and $\gamma$ close to the critical values of $U/W$ are directly associated with the dielectric results reported in Ref.\,\cite{Dressel} and also with an entropy enhancement in that region. A pronounced electrocaloric effect should be expected under such conditions \cite{ECE}.
The results discussed so far are key in understanding the enhancement of $\Gamma_\varepsilon$ in the immediate vicinity of the critical end point due to the phases coexistence.
When dealing with a polarizable medium, the charges tend to be screened by opposite sign charges in a way that the total charge is reduced and, as a consequence, the Coulomb potential is reduced as well. By analyzing the behavior of $\varepsilon_1$ reported in Ref.\,\cite{Dressel} it is observed that upon pressurization $U/W$ is reduced, the system becomes mainly metallic and $\varepsilon_1$ is drastically reduced. Thus, as a consequence of the lowering of $\varepsilon_1$, the distance between charges must also be reduced. Such a lowering of the distance between charges must increase screening effects in order to weaken the Coulomb repulsion. As a consequence, the \emph{electronic} isothermal compressibility \cite{Kotliar,Hassan}, $\kappa_T = -\frac{1}{v}\left(\frac{dv}{dp}\right)_T$, is enhanced for  $(U/W) = (U/W)_{crit}$, cf.\,Fig.\,\ref{Fig-2} b). Next, we discuss the realization of an electronic Griffiths-like phase close to the first-order transition line. Given the intrinsic spatial disorder and randomness inherent to the coexistence region due to the competition between metallic and insulating phases, we make an association with an electronic Griffiths-like phase \cite{Griffiths1969, Vojta2006a,Vojta2006b,Magen2006,Vojta2006b}.
The canonical Griffiths phase is defined as rare ferromagnetic regions embedded in a lattice with inserted quenched disorder upon randomly removing some of the spins from the lattice, which in turn affects the critical temperature of the paramagnetic to ferromagnetic transition, cf.\,Refs.\,\cite{Griffiths1969, Vojta2006b, Vojta2013}. This leads to a probability $p'$ of having a vacancy in the lattice and a probability ($1 - p'$) of having a spin. Even if there are many vacancies in the system, finite-size ferromagnetic regions can still be present. If the percolation threshold is achieved, i.e., the number of vacancies achieve a threshold, the ferromagnetic transition is suppressed. In an analogous way, when the system is a Mott-insulator and external pressure is applied, it is tuned into the coexistence region composed by metal and Mott-insulator phases and thus metallic non-polarized puddles start to randomly emerge into the system as pressure is increased. Similar to the magnetic Griffiths phase, we can associate a probability $p'$ of having a non-polarized region (metallic puddles) and a probability of ($1 - p'$) of finding a polarized region (ferroelectric Mott insulator). If pressure is continuously applied through the metallic spinodal line, the percolation threshold is achieved and the system becomes fully metallic, i.e., the ferroelectric phase is suppressed analogously to the ferromagnetic one in the case of the magnetic Griffiths phase. In other words, the metallic puddles play an analogous role as the vacancies (non-magnetized regions) in the magnetic Griffiths phase. Also, as the number of vacancies in the Griffiths phase increases, the system becomes more diluted and the critical temperature of the ferromagnetic transition is shifted to lower temperatures. At the percolation threshold, the critical temperature vanishes since long-range magnetic ordering is no longer attained. Analogously, the dielectric constant measurements for the spin-liquid candidate reported in Ref.\,\cite{pustgow2} show that the temperature associated with the maximum response of the dielectric constant is shifted to lower temperatures upon increasing pressure. Indeed, the application of pressure makes the system to approach the phases coexistence region and thus metallic puddles start to emerge into the insulating matrix. Hence, the insulating phase becomes diluted due to the presence of such non-polarized regions and, as a consequence, the temperature associated with the maximum of the dielectric response is shifted to lower temperatures. Thus, in a direct analogy with the magnetic Griffiths phase \cite{Vojta2013}, the relaxation-time is enhanced upon crossing the first-order transition line because the time-scale to reverse the electric polarization of the polarized regions is higher when compared with the case of the fully insulating phase. Furthermore, the magnetic Griffiths phase can be seen either as rare ferromagnetic regions within a paramagnetic (vacancies) bulk system, namely the ferromagnetic Griffiths phase, or rare vacancy-rich regions embedded in a ferromagnetic bulk system, being in this case a  paramagnetic Griffiths phase, cf. discussed in Ref.\,\cite{Vojta2013}. Analogously, in the phases coexistence region of the Mott transition, we have either polarized insulating rare regions immersed in a non-polarized (metal) bulk system or the opposite, i..e.,  non-polarized rare regions embedded into an insulating matrix, cf. lower and upper insets of Fig.\,\ref{Fig-1}.
Hence, we identify the so-called rare regions with the metallic and insulating close to the critical point, cf.\,lower inset of Fig.\,\ref{Fig-1}.
The analogy between the canonical Griffiths phase \cite{Griffiths1969} and the Mott coexistence region lies in the fact that when crossing the first-order transition line ($x$ = 0.5) polarized puddles coexist with a non-polarized metallic matrix or vice-versa, in the same way as ferromagnetic clusters are immersed in a non-magnetic medium in the \emph{classical} Griffiths phase.
Considering the critical fluctuations, we are dealing with dynamic rare region effects \cite{Vojta2006b}. In this case, following discussions made in Ref.\,\cite{Vojta2006b}, the \emph{puddles} (rare region) \emph{auto}correlation function $C_{RR}(t)$, reads:
\begin{equation}
C_{RR}(t) \sim \int_0^\infty dL_{RR}w(L_{RR})exp[-t/\tau(L_{RR},r_0)],
\label{acf}
\end{equation}
where $w(L_{RR})$ refers to the probability of finding a puddle of size $L_{RR}$ and $\tau(L_{RR},r_0)$ is the relaxation time of the considered  \emph{puddle} in a reduced temperature $r_0$.  Interestingly, Eq.\,\ref{acf} embodies the key parameters associated with the coexistence region close to the Mott transition. 
It turns out that $\tau(L_{RR},r_0)$ follows a similar behavior as the relaxation time proposed in Ref.\,\cite{Casalini2}. 
Application of pressure affects the size and dynamics of the rare regions (puddles). The interplay between metallic and insulating coexisting phase is governed by $x$, which is dictated by the Hubbard parameter. It is well-established in the literature that the presence of disorder gives rise to broadening effects in the response function. In the present case, because of the spatial randomness of the  metallic/insulating puddles it is tempting to state that the Mott transition would be continuous and thus of a second-order character  \cite{Senthil2008,Mishmash2015}, as reported for another molecular conductor \cite{Itou2017}. The phases coexistence region close to the first-order transition line of any system might be treated
as a region comprising two distinct fluids, so that upon crossing the first-order transition line we could assign the transition as a liquid-liquid-type transition \cite{Debenedetti}. Given the slow dynamics intrinsic to the coexistence region \cite{Dressel,Kundu}, our findings are suggestive of a diverging effective mass upon approaching the critical Hubbard parameter  \cite{Senthil2008}. 

\section{Conclusion}
Our analysis of the phases coexistence region, based on a compressed cell $S = 1/2$ Ising-like model, is universal. The here proposed electric Gr\"uneisen parameter quantifies the electrocaloric effect. Using scaling arguments, we have shown that close to \emph{any} critical point the relaxation time is entropy-dependent. This feature might be relevant in the optimization of caloric effects close to critical points \cite{ECE}. The coexistence region of the Mott transition can be identified as an electronic Griffiths-like phase. Our findings suggest that a Griffiths-like phase can emerge in any system within a coexistence region presenting both slow dynamics and randomness
\cite{ghosh,dagotto,Griffiths1979}.

\section*{Supplementary Material}
See Supplementary Material for a more detailed discussion about the mathematical function used to describe the metallic volume fraction $x$. The dielectric Gr\"uneisen parameter, pressure-induced effects on the dielectric response, and a broader analysis of the intrinsic slow-dynamics within the Griffiths phase are also presented.

\begin{acknowledgments}
We acknowledge financial support from the S\~ao Paulo Research Foundation -- Fapesp (Grants 2011/22050-4); National Council of Technological and Scientific Development -- CNPq (Grants 302498/2017-6 and 305668/2018-8); Capes - Finance Code 001 (Ph.D.\,fellowship of L.S.\,and I.F.M.).

IFM and LS contributed equally to this work.
\end{acknowledgments}


\section*{Data Availability Statement}
The data that support the findings of this study are available from the corresponding author upon reasonable request.


\end{document}